\documentclass[a4paper,12pt]{article}

\usepackage{amsmath,amssymb,amscd}
\usepackage{graphicx}
\usepackage{hyperref}

\ifcsname shortitle\endcsname
  \else%
  \fi%

\providecommand{\shorttitle}[1]{} 
\providecommand{\shortauthorlist}[1]{} 
\providecommand{\name}[1]{#1} 
\providecommand{\address}[1]{\\{}#1} 
\providecommand{\and}{\\{} and } 

\providecommand{\url}[1]{\texttt{#1}} 
\providecommand{\href}[2]{#2} 

\newcommand{\tsedoi}[1]{\url{http://dx.doi.org/#1}}
\newcommand{\tsedoiextra}[1]{} 

\newcommand{\tsevec}[1]{\mathbf{#1}}
\newcommand{\tsemat}[1]{{\mathbf{\textsf{#1}}}}
\newcommand{\tsematg}[1]{{\boldsymbol{#1}}}

\newcommand{\xvec}{\tsevec{x}}

\newcommand{\unitvec}{\mathbf{1}} 

\newcommand{\Bmat}{\tsemat{B}}
\newcommand{\Dmat}{\tsemat{D}}
\newcommand{\Gmat}{\tsemat{G}}
\newcommand{\Jmat}{\tsemat{J}}
\newcommand{\Mmat}{\tsemat{M}}
\newcommand{\Smat}{\tsemat{S}}
\newcommand{\Umat}{\tsemat{U}}
\newcommand{\Xmat}{\tsemat{X}}

\newcommand{\Sigmamat}{\tsematg{\Sigma}}

\newcommand{\unitmat}{\hbox{\textsf{1}\kern-.25em{\textsf{I}}}}

\newcommand{\Gcal}{\mathcal{G}}

\newcommand{\Mbb}{\mathbb{M}}

\newcommand{\Dreduced}{{\hat{D}}}

\newcommand{\Trans}{\mathrm{T}}

\begin{document}

\title{Embedding graphs in Lorentzian spacetime}
\shorttitle{Embedding graphs in Lorentzian spacetime} 
\shortauthorlist{J.R.~Clough, T.S.~Evans} 

\author{
\name{James~R. Clough, Tim~S. Evans}
\address{Centre for Complexity Science, Imperial College London,\\ South Kensington campus, London, SW7 2AZ, United Kingdom}
}
\maketitle


\begin{abstract}
Geometric approaches to network analysis combine simply defined models with great descriptive power.
In this work we provide a method for embedding directed acyclic graphs into Minkowski spacetime using Multidimensional scaling (MDS).\newline
First we generalise the classical MDS algorithm, defined only for metrics with a Euclidean signature, to manifolds of any metric signature.
We then use this general method to develop an algorithm to be used on networks which have causal structure allowing them to be embedded in Lorentzian manifolds.
The method is demonstrated by calculating embeddings for both causal sets and citation networks in Minkowski spacetime.
We finally suggest a number of applications in citation analysis such as paper recommendation, identifying missing citations and fitting citation models to data using this geometric approach.
\end{abstract}

\section{Introduction}
One success of network science has been to identify that some complex systems can be simplified by considering just the topology of the pairwise interactions between their parts.
Abstracting a complex system as a graph can bring physical insights and predictive power.
Yet these graphs can still be very complicated. 
Network geometry is an approach which further abstracts the system by modelling the nodes of the network as points in a geometric space.

Most existing approaches use Riemannian spaces, the simplest example of which is Euclidean space.
Random Geometric Graphs (RGG) are graphs embedded in Euclidean space \cite{Dall2002, Penrose2003, Xie2015}.

Recently there has been much interest in geometric approaches to the study of networks in  non-Euclidean spaces \cite{Krioukov2010, Krioukov2012, Clough2014, Wu2015}.
Embedding in hyperbolic spaces can yield scale free, clustered networks with community structure illustrating the remarkable power that geometric approaches have to recover complex network properties.

A well established geometric approach to data analysis is \emph{Multidimensional Scaling} (MDS), a technique to give data expressed as distances or similarities a spatial representation \cite{Cox2000}.
In most MDS analysis, the space used for that spatial representation has been Euclidean, and the technique, as usually described, requires a Riemannian manifold, where the triangle inequality is maintained.

MDS has been used in network science, to fit models of RGGs to networks from real data, for example, from protein interactions \cite{Higham2008}.
Normally, the MDS algorithm takes as an argument pairwise distances between objects, so when applying it to simple networks, where only binary pairwise relations exist, these distances have to be inferred from the network structure.
In the simplest Euclidean case (as in \cite{Higham2008}), the shortest path on the network is used as an estimate for the distances between vertices, from which MDS is used to calculate coordinates.
Once these coordinates have been calculated, a new RGG can be built from them, and if it is similar to the original graph, the initial geometric assumption was a good one.

In this paper we will consider networks where each node is associated with a particular time and directed edges between nodes represent causal relations. 
Such a network forms a \textbf{directed acyclic graph} (DAG)\cite{Clough2013}. 
Instead of embedding a network in geometric space alone, the causal ordering of the nodes in a DAG suggests that an embedding in space and time is needed.  
The causal structure of such a network has the same constraints as the causal structure of spacetime as used in special and general relativity \cite{Hawking1973, Brightwell2015}.\newline
This suggests that the geometries used in relativity, which are pseudo-Riemannian are the appropriate ones to use because of the special properties of a time dimension. 
In particular, we will consider \emph{Lorentzian spacetimes}, a special case of pseudo-Riemannian manifolds in which there is one time dimension with some number of spatial dimensions.

Euclidean space, being flat and isotropic, is the simplest Riemannian manifold.
Analogously, flat isotropic spacetime is Minkowski spacetime, the simplest Lorentzian manifold.
In this paper, we first generalise classical MDS to allow converting distances into coordinates for pseudo-Riemannian manifolds.
We then show how this allows embedding of DAGs in Minkowski spacetime\footnote{Somewhat confusingly, in some of the literature on MDS (particularly in psychology - see for instance Shepard \cite{Shepard1964}), a Minkowski metric refers to a distance measure of the form $d_{xy} = \left[ \sum_i (x_i - y_i)^\lambda \right] ^{1/\lambda}$. 
This is \emph{not} what we mean by a Minkowski metric in this paper, rather we mean the Minkowski spacetime of special relativity\cite{Minkowski1910}.}, 
and that the coordinates of geometric graphs in Minkowski spacetime can be successfully recovered.
We then illustrate this technique by finding coordinates for some citation networks, which naturally form DAGs and suggest applications such as paper recommendation.

\section{Lorentzian Multidimensional Scaling}

\subsection{Review of Standard MDS}

We will begin by briefly summarising the details of standard MDS in Euclidean space.
Suppose we have $N$ objects, and are given the squared Euclidean distance, $S_{ij}$ between each pair.
We wish to find the co-ordinates of the objects, which will be $D$ dimensional vectors, $\mathbf{x}_i$ for each object $i$, such that they fit the constraint that $|\mathbf{x}_i - \mathbf{x}_j|^2 = S_{ij}$.

The classical MDS algorithm solves this problem by using this $N \times N$ matrix of square distances, $\Smat $, and then constructing the double centred matrix $\Bmat = - \frac{1}{2}\Jmat\Smat\Jmat$ where
$ \Jmat = \unitmat - \frac{1}{N} {\unitvec}.{\unitvec}^{\Trans} $.

It can then be shown (details are available in \cite{Cox2000}) that
\begin{equation}
 \Bmat = \Xmat \Xmat^\Trans
\end{equation}
where $\Xmat$ is an $N \times D$ matrix of co-ordinate vectors $\xvec$ which satisfy the constraint of recovering the original distances, and with the centre of mass of the coordinates at the origin.
$\Bmat$ is guaranteed to be semi-positive definite (i.e.\ it has no negative eigenvalues).
So we can then find (up to a distance-preserving symmetry transformation) the coordinates in $\Xmat$ by decomposing $\Bmat$ into
\begin{equation}
\Bmat = \Umat \Sigmamat \Umat^\Trans
\end{equation}
where $\Sigmamat$ is a diagonal matrix of the eigenvalues of $\Bmat$, and $\Umat$ a matrix of its eigenvectors. A solution is given by
\begin{equation}
\Xmat = \sqrt{\Sigmamat} \Umat
\end{equation}
This process yields coordinates in $N$ dimensions, but only $D$ of the eigenvalues will be non-zero.
It is possible retrieve coordinates in fewer dimensions, by using only the largest $\Dreduced$ eigenvalues and their corresponding eigenvectors.
The larger eigenvalues correspond to principle components, meaning that using them as the coordinates minimises the square difference between the original distances we started with, and the ones calculated from these inferred coordinates.
These coordinates are in this sense the most accurate $\Dreduced$ dimensional representation of the original data.

\subsection{Lorentzian MDS}
Minkowski spacetime is a combination of a $d$-dimensional Euclidean space, and one time dimension forming a $(d+1)$-dimensional manifold.
A point $i$ in this space, has coordinates $\mathbf{x}_i$ consisting of a time coordinate, $x_i^0$, and spatial coordinates $x_i^k$, with $k=1, 2, ..., d$.
The Minkowski separation between two such spacetime points $i$ and $j$ is given by
\begin{equation}
M_{ij} = M(\mathbf{x}_i, \mathbf{x}_j) =
-(x_i^0 - x_j^0)^2 + \sum_{k=1}^{d} (x_i^k - x_j^k)^2
\label{M_Metric}
\end{equation}
Pairs of points can then be classified into three types:
for a positive separation the pair is spacelike separated,
for a negative separation the pair is timelike separated,
while exactly zero separation means the pair is lightlike separated.
We can now ask the same question that classical Euclidean MDS poses: given pairwise separations $M_{ij}$, for points in this space, can we recover coordinates which respect these separations?

Proceeding with the classical Euclidean algorithm we can construct the double centred matrix $\Bmat$ as before using $\Bmat = - \frac{1}{2} \Jmat \Mmat\Jmat$. 
However we now encounter a problem when decomposing $\Bmat$.
Previously the eigenvalues were guaranteed to be non-negative, but now we find one negative eigenvalue corresponding to the time dimension's negative sign in equation~\ref{M_Metric}. 
Since we need to take the square root of these eigenvalues, and we want real coordinates this is a problem.

It turns out that the changes required to the classical MDS algorithm are remarkably simple (details are given in appendix A).
Instead of looking for a matrix of coordinates $\Xmat$ such that $\Bmat= \Xmat \Xmat^\Trans$, we now search solutions to
\begin{equation}
 \Bmat = \Xmat \Gmat \Xmat^\Trans
\end{equation}
where $\Gmat$ is matrix representing the metric of the embedding space. 
For traditional MDS with its Euclidean space $\Gmat$ is just the identity matrix so this factor drops out from the analysis.
However for DAGs where nodes are also associated with a time coordinate, we choose $\Gmat$ to represent the Minkowski metric. 
In our conventions, this is a diagonal matrix with $-1$ in the first column and $+1$ in the others.
We again decompose $\Bmat$ and now need solutions to
\begin{equation}
\Xmat\Gmat\Xmat^\Trans = \Umat \Sigmamat \Umat^\Trans \,.
\end{equation}
The difference to traditional MDS is that the $-1$ present in $\Gmat$ changes the sign of the one negative eigenvalue in $\Sigma$.  
This allows us to take the square root of $\Sigmamat$ as we do in classical Euclidean MDS.

\subsection{Causal Set DAGs and Minkowski spacetime}
To use Lorentzian MDS on networks, we require a method of estimating the separations between every pair of nodes from the network structure.
We will do this using ideas from causal set theory.
A causal set is a locally finite partially ordered set.

In the causal set approach to quantum gravity, the underlying structure of spacetime is postulated to be a causal set.
Spacetime is seen as discrete at the Planck scale, and the continuous spacetime we perceive emerges at larger length scales.\cite{Bombelli1987, Bombelli1989, Dowker2006}.
The only structure present are elements of the set and the relation, $\prec$ between pairs of elements.
In the correspondence between the discrete causal set, and the continuous spacetime that emerges, the relation $\prec$ corresponds to timelike separation where $x \prec y$ corresponds to $x$ being in the past of $y$, and spacelike separated pairs are not related.
Causal sets give us a natural way of discretising spacetime with related elements, and we can use them in our approach because they have the same structure as a DAG.
In physics a timelike separated pair are allowed to be causally related, as information could pass from one to the other, from past to future.
However a spacelike separated pair can never be causally related as light could not reach one event from the other.
This is why we will associate timelike separation in the continuous spacetime with a link representing a causal relation in the network.\newline

We can now construct the equivalent of an RGG in Minkowski space.
We begin by assigning coordinates in $\Mbb^D$ to $N$ points, by sampling uniformly at random\footnote{The range of coordinates is chosen for simplicity but this does not need to be the case in general.} from $[0, 1]^D$.
We will denote the coordinates of point $i$ by $x_i^\mu$, $\mu=0,1,\ldots,d$, where $\mu=0$ is the time coordinate.
We then construct the causal set graph, $\Gcal_D$ by placing an edge from $i$ to $j$ if $M_{ij} < 0$ and directing that edge from the past to the future.
The fact that edges in the graph now represent causal relations illustrates why the graph is necessarily a DAG, as closed causal loops are forbidden.\newline

\begin{figure}
\includegraphics[width=1.0\textwidth]{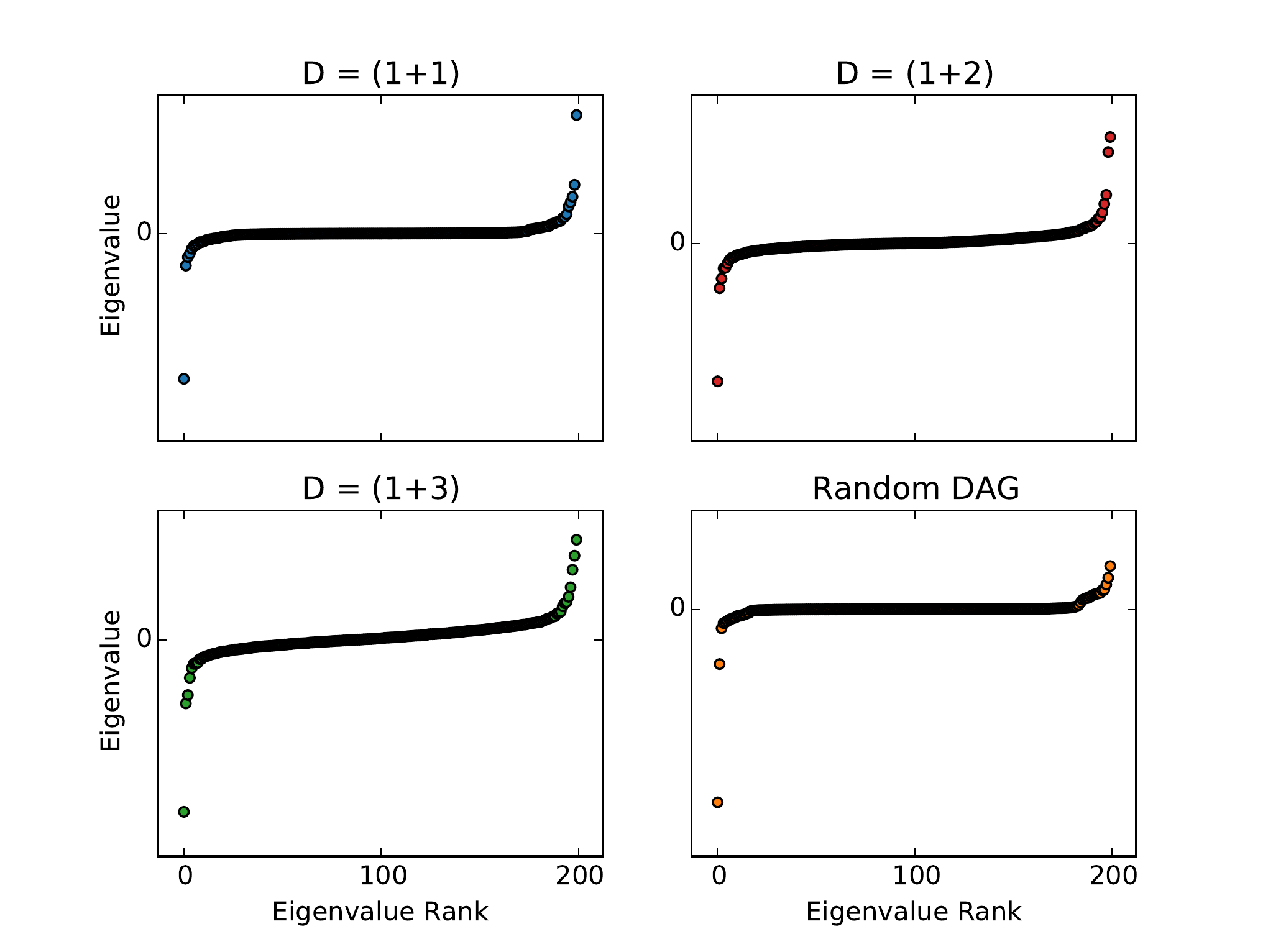}
\caption{The eigenvalue distribution for MDS of 200 point causal sets in $1+1$, $1+2$, and $1+3$ dimensions, and for a random DAG.
In all cases, one large negative eigenvalue is seen, corresponding to the one timelike dimension.
For the causal sets, this timelike dimension is the one time dimension in the Minkowski spacetime they are embedded in.
For the random DAG, this corresponds to the time ordering which could be created as a consequence of the acyclic property of this graph.
We observe $d$ large positive eigenvalues corresponding to the spatial dimensions in each causal set, illustrating that the coarse graining of the causal set does not prevent the MDS algorithm from successfully identifying the principle components of the space.
For a given number of points, higher dimensionality will mean fewer relations between the causal set's elements and since these relations are the information used by the MDS algorithm its ability to cleanly pick out $d$ dimensions diminishes as $d$ increases.
However for the random DAG there is no clear separation of large eigenvalues suggesting there is no natural dimension to an embedding Minkowski spacetime.}
\label{fig:eigenval}
\end{figure}

Figure~\ref{fig:eigenval} shows the distribution of the eigenvalues $\Sigmamat$ for causal sets and random DAGs\footnote{By a random DAG we mean an Erd\H{o}s-R\'enyi random graph, with the nodes placed in a random order and edges directed along that order, and then the graph transitively completed. The example in the figure uses an Erd\H{o}s-R\'enyi graph with an average degree of 25 but the result is general.}.

\subsection{Embedding Graphs using Lorentzian MDS}
To use MDS on a network we must estimate the separation matrix $\Smat$ using the network's structure.
For Euclidean MDS, the separation is always a non-negative number and the shortest path between nodes in the graph is a natural and effective estimator for the distance.
However in Minkowski spacetime, the separation of points is not always positive, so the the number of steps along some path is not going to be measure of all Lorenztian separations. 
The solution is to estimate spacelike and timelike separations separately when studying a DAG.

Suppose we have two connected nodes in the graph, meaning they are timelike separated.
It was conjectured in \cite{Myrheim1978} and later shown in \cite{Bollobas1991} that for timelike separated points $i$ and $j$ in $\Gcal_D$ the length of \emph{longest path} $L_{ij}$ between two connected nodes, say $i$ and $j$, is proportional their timelike separation (in the limit of long separations, and where the longest path must respect the edge's direction).
So in this case we set $S_{ij} = -1 \times L_{ij}^2$.

\begin{figure}[htb]
\includegraphics[width=1.0\textwidth]{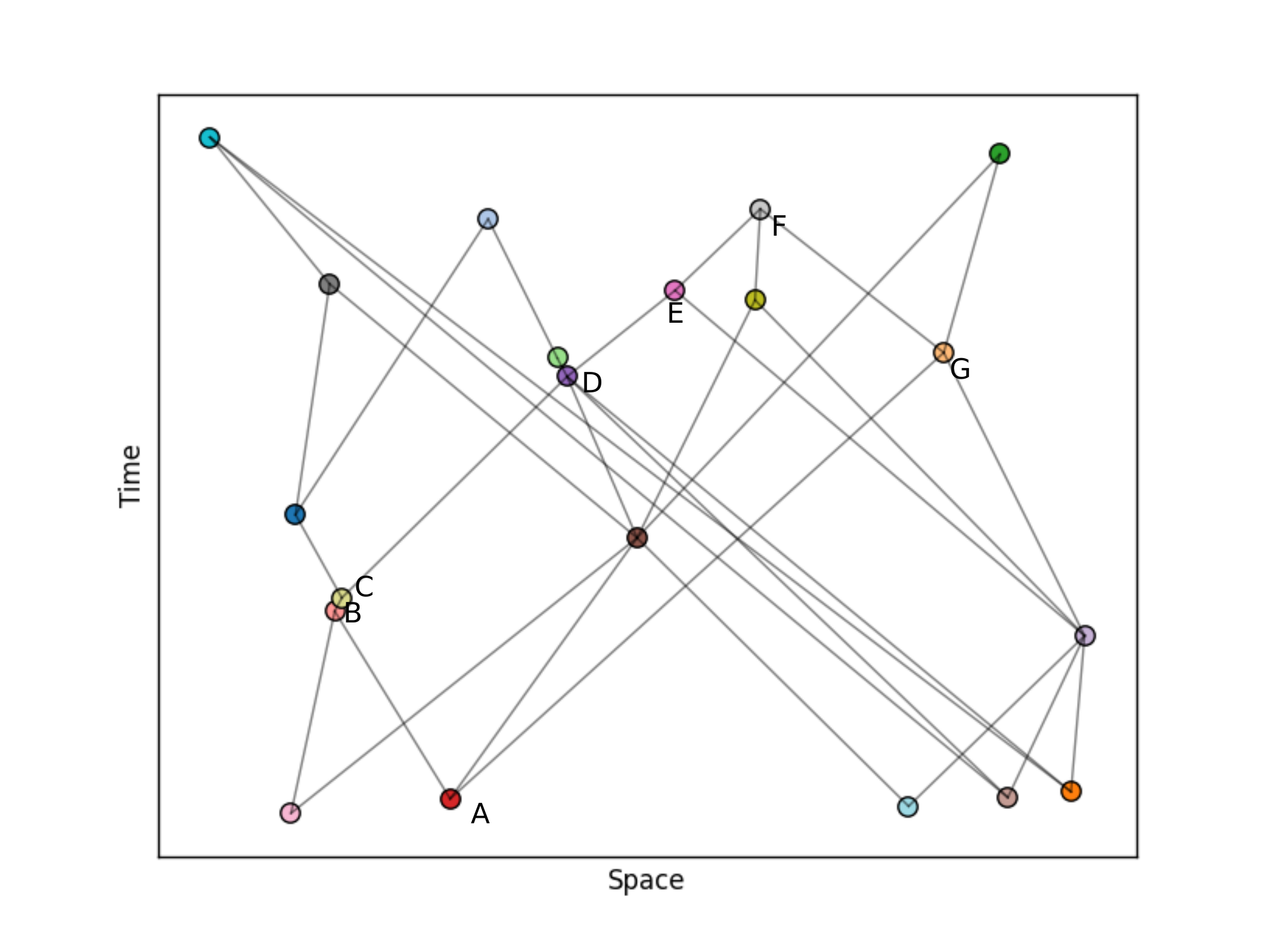}
\caption{The timelike separation between nodes A and F is approximated as 5 units - as this is the number of edges in the longest direction-respecting path between them.
Nodes B and G are spacelike separated.
To estimate this separation we find a pair of points in their mutual past and future.
In this case, the only such pair is (A, F).
The naive spatial separation between (B, G) is then given by the timelike separation between (A, F) so is also in this case 5 units.
Note, only the edges not implied by transitivity have been drawn.}
\label{fig:causal_set}
\end{figure}

Finding the distance between spacelike pairs is more challenging and to our knowledge there is no solution as easily calculated as the longest path is for timelike pairs.
Good approximations are known however, and we will use a very simple one, described in \cite{Brightwell1991,Rideout2009a} as `naive spatial distance'.
Suppose we have two disconnected points $i$ and $j$ in the $\mathcal{G}_D$ meaning they are spacelike separated.
We then look for a pair of nodes, $k$  and $l$, where $k$ is in the future of both nodes $i$ and $j$ while $l$ is in the past of both $i$ and $j$\footnote{If no such pair exists, we set the spacelike distance of $i$ and $j$ equal to some maximal distance which is a parameter of the algorithm. In the examples shown here, we used the length of the longest path in the graph as this parameter.}. 
We then chose nodes $k$ and $l$ so as to minimise the length of the longest path between $k$ and $l$ (which are necessarily connected, via paths through $x$ and $y$). 
The timelike separation between $k$ and $l$ is then used as an estimate for the spacelike separation between $i$ and $j$.
Figure~\ref{fig:causal_set} shows an example of how we estimate timelike and spacelike separations.

This estimate is simple and at first appealing, but fails in more than two dimensions for large graphs (hence the `naive' in the name). 
Nonetheless we find it is sufficient for our purposes\footnote{We tried using the two-link method described in \cite{Rideout2009a} but found that for small graphs with fewer than $500$ nodes too many cases had no two-links and so the spacelike separation couldn't be calculated}. 
This is partly because it is inaccurate only for large causal sets but also because in the MDS algorithm each point's coordinates is fixed by many distances, both timelike and spacelike which limits the effect of noise some poor estimations of spacelike separations.
We can then set $S_{ij} = L_{kl}^2$ for the chosen $k$ and $l$.

These timelike and spacelike distances define our separation matrix $\Smat$ (where timelike separation has the $-$ sign in our conventions). 
Finally we use the algorithm described in the previous section to assign coordinates in some $\Dreduced$-dimensional Minkowski spacetime $\Mbb^\Dreduced$ to each vertex in the DAG.

\section{Testing Lorentzian MDS on Causal Sets}
Given a causal set graph, $\Gcal_D$ it is possible, in principle and for large enough $N=|\Gcal_D|$, to recover all properties of the spacetime (up to a factor of the density of the sprinkling)\cite{Hawking1973, Dowker2004}.

In Minkowski spacetime there is only one parameter to recover, $D$, and this can be estimated for the process described above and for DAGs in general \cite{Myrheim1978, Reid2003, Clough2014}.
Our task here is to recover not properties of the manifold in which the nodes are embedded, but to find the full details of that embedding.  

If the graph was originally made sprinkling points into $\Mbb^D$ then we know that an exact embedding is possible (since the original sprinkled coordinates must be a solution), and so the embedding algorithm should approximately recover the original coordinates (up to distance preserving factors).
If the graph was not, then we may only be able to find an approximate embedding.\newline

\begin{figure}
\includegraphics[width=1.0\textwidth]{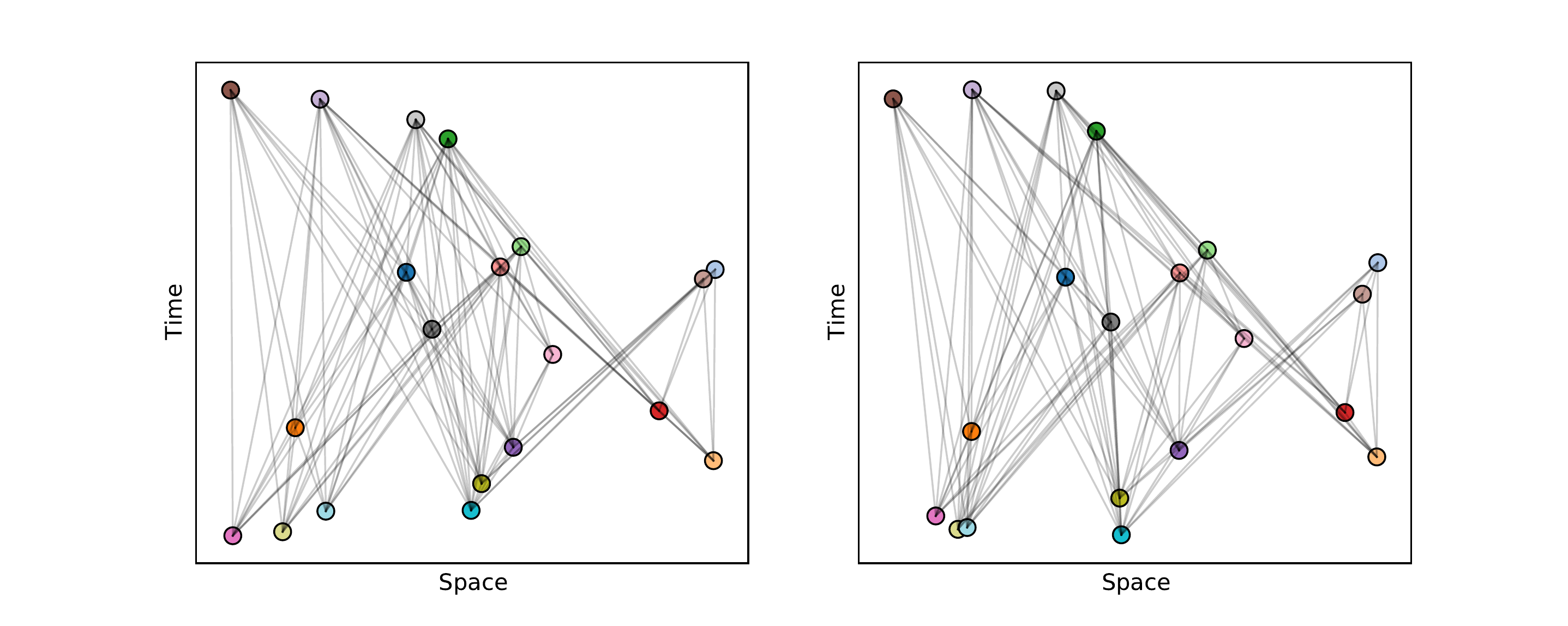}
\caption{20 points of a 200 point causal set, (left) and its embedding after the MDS algorithm.
The similarity between the two plots shows the success of the embedding algorithm in finding the point's coordinates using only the edges of the graph.}
\label{fig:M_MDS}
\end{figure}

As is the case for classical MDS, our Lorentzian MDS is guaranteed to recover the coordinates of points when the exact distances are used as the algorithms input.
However, when distances are estimated using graph topology, the pairwise separations will be noisy as they are coarse grained by the discrete graph (see figure~\ref{fig:M_MDS}).
To assess the reliability of this algorithm we will test it first on the causal set model described above.  
We will take the coordinates produced by the Lorentzian MDS algorithm use them to rebuild the graph by again placing edges only between timelike separated pairs.
If the overlap between the edges between nodes in the recreated graph, and in the original graph is high, the embedding is an accurate one, and similarly if the overlap is low the embedding is poor.
As in \cite{Higham2008} we will measure this using the sensitivity (the fraction of the correct edges which were predicted) and specificity (the fraction of correct non-edges which were predicted).\newline

\section{Citation Networks in Minkowski spacetime}
\begin{figure}
\includegraphics[width=1.0\textwidth]{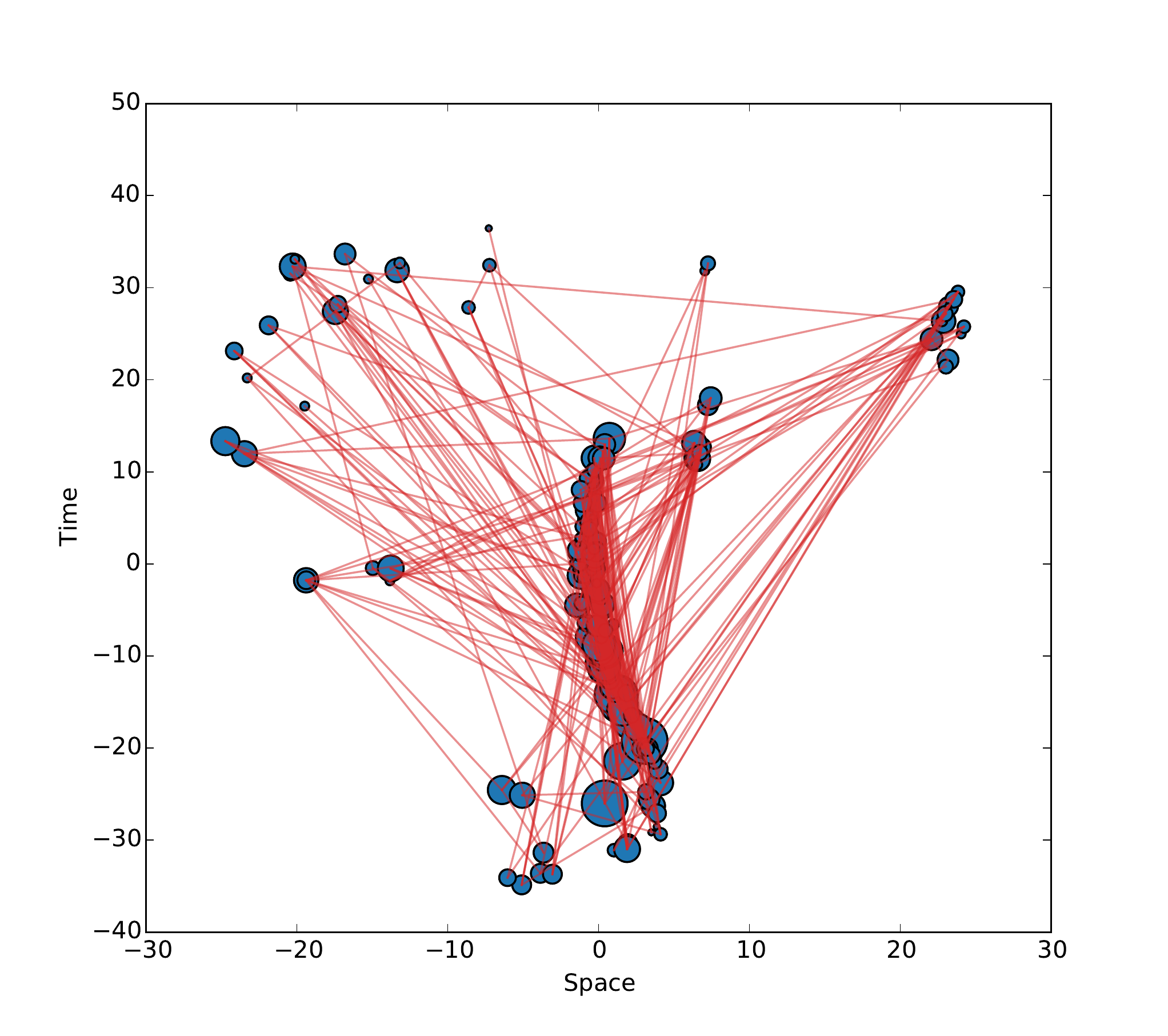}
\caption{An $D=1+1$ embedding of the top $200$ most cited papers in the \texttt{hep-th} citation network.
Node size is proportional to the number of citations, and lines correspond to citations amongst these papers.
A large group of papers is visible in the middle of the plot, forming a long chain of citations, as well as some more isolated papers on either side.
A small number of spacelike citations are visible (those edges more than $45\deg$ from vertical) because this two-dimensional embedding is not perfect, but only the optimal set of coordinates found by the MDS algorithm.}
\label{fig:hep-th_200}
\end{figure}
\begin{figure}
\includegraphics[width=1.0\textwidth]{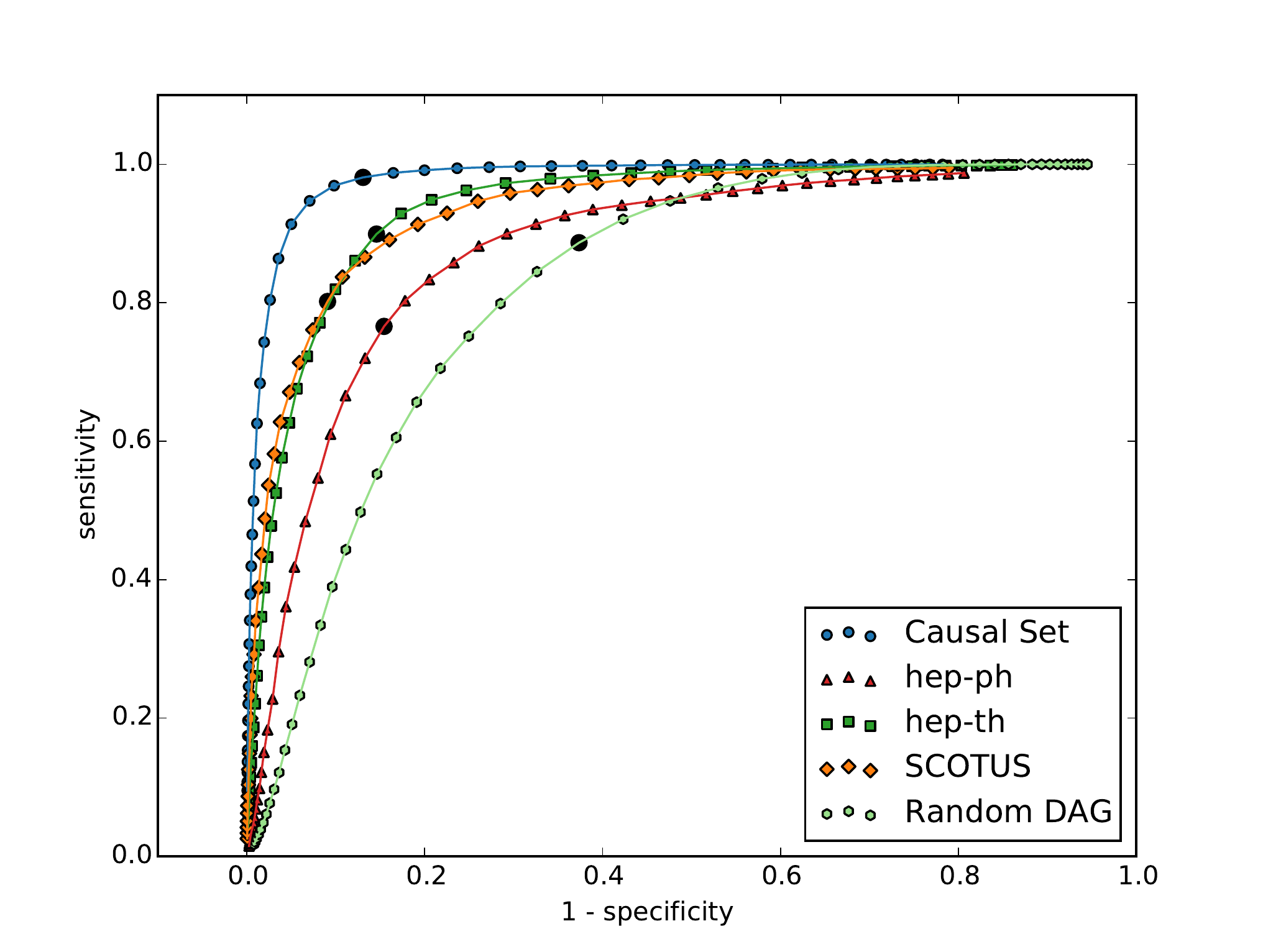}
\caption{Sensitivity and specificity are measured for various $c$ values for $D=(2+1)$ dimensional embeddings of $5$ networks.
Where the curve reaches the top-left of the plot we are in the $c\approx1$ regime ($c=1$ denoted by the large black point on each curve) where the trade-off between false positives and false negatives is balanced.
The shape of these ROC curves measures the effectiveness of the embedding and it is clear that the the causal set performs best, the random DAG worst and the citation networks in between.}
\label{fig:auc_example}
\end{figure}

Our MDS algorithm is able to find accurate embeddings for randomly sprinkled causal sets.
We can now attempt to embed networks formed from real social systems, and here we will use citation networks from the arXiv (2003 KDD cup datasets) and the US Supreme Court\cite{Fowler2008}, as well as random DAGs for comparison.
Recall that the dimensionality of the embedding is something we can choose, by selecting the $\Dreduced$ largest eigenvalues in the MDS algorithm.
To measure the effectiveness of the embedding we will compare the original network, with a new network generated from the coordinates determined by the MDS algorithm.
We want to measure the effectiveness of a classifier which predicts edges in the network from the MDS coordinates.
To do this we will use the established method of the area under the receiver-operator curve, AUC.
Varying a continuous parameter, the sensitivity and specificity of the embedding is measured, and plotted, as in figure ~\ref{fig:auc_example} and the area under this curve describes the quality of the classifier.
The continuous parameter we will vary is the speed of light (or the speed information can be transferred) in the Minkowski space, $c$.
Previously, we have set $c=1$, but varying this speed will change which nodes are connected in new network generated from the MDS coordinates.
Now, nodes $i$ and $j$ are connected if their coordinates satisfy
\begin{equation}
-c(x_i^0 - x_j^0)^2 + \sum_{k=1}^{d} (x_i^k - x_j^k)^2 < 0
\end{equation}
For small values of $c$, very few nodes are connected and so the specificity is high (few false positives) but the sensitivity is low (many false negatives).
For large values of $c$, many nodes are connected and so the reverse is true.

We will measure the ease of embedding a network by taking the mean of the area under this curve for networks of size $100$-$500$.
In the case of the citation networks this was done by randomly sampling intervals of this size from the citation network. 
In the cases of the causal set graph, and the random DAGs, many instances of the required sizes are stochastically generated.

\begin{figure}
\includegraphics[width=1.0\textwidth]{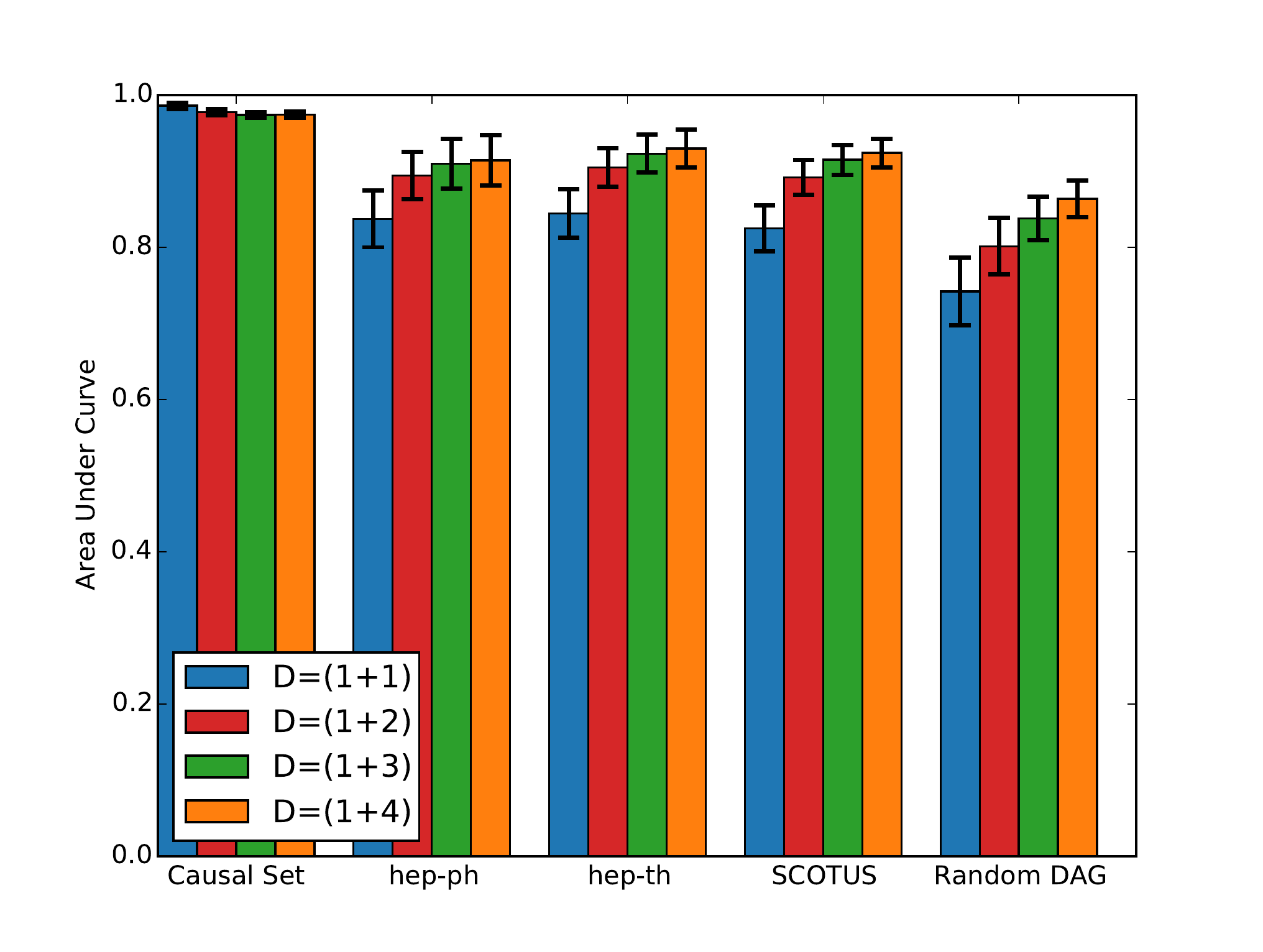}
\caption{The AUC for 5 types of DAG, embedding in Minkowski spacetime of various dimensions.
In this plot, the causal set graphs are created from sprinkling uniformly into a $(d+1)$-dimensional spacetime, then embedded back into that same dimensional spacetime.
Their AUC values therefore represent the ideal case, where we know an embedding is possible.
The random DAGs show the worst embeddings.
Their AUC values represent the success of an embedding for a graph which has no structure.
They are noticeably more embeddable in higher dimensions which is because there are more degrees of freedom in which to assign coordinates while maintaining the randomly placed links.
The bars show means and standard deviations of the AUC for random DAGs of size $N=100$ to $N=500$.
Citation networks from the arXiv, and the US Supreme Court fall between these extremes, illustrating the presence of some structure which makes them easier to embed in spacetime.
We sampled $250$ intervals with between $100$ and $500$ nodes randomly from each citation network, and the bars show means and standard deviations for the AUC in each case.}
\label{fig:embed_auc}
\end{figure}

\section{Discussion and Applications}
Finding a good geometric embedding of a network provides a powerful tool for the analysis of that network as it allows standard geometric techniques and intuition to be used.
Calculations of network properties can be made more efficient, for example, when finding optimal routes from one node to another, the node coordinates provide local information which can improve routing algorithms\cite{Krioukov2009}.
Coordinates resulting from geometric embedding also provide a natural visualisation for a network.
Such visualisations are used in bibliometrics to help identify distinct fields or assist literature reviews\cite{Eck2014}.\newline

In the cases of citation analysis, where we conjecture that the spatial dimensions that result from a geometric approach correspond to similarity in the topic of a paper, our approach yields spatial similarities between papers while accounting for the time difference in their publication.
Once embedding coordinates are known, the idea that nodes may be `similar' can be expressed as nodes being close by using an appropriate metric, which need not be the Lorentzian one we used in the construction of the coordinates. 
Two nodes might be spacelike separated, so have no direct links, yet be close in a Euclidean sense because MDS calculates coordinates globally using information from all vertices and edges.
Papers could be recommended if spatially close to others a reader has interest in even if there are no local connections between them, potentially bringing work or authors to the attention of readers who are not aware of them.
Spatial similarity can also be used to define clusters. 
The idea of `centrality' or importance of a node has a natural representation in terms of the density of points in the geometric neighbourhood of the point linked with the chosen node.\newline
  
Another use of this approach is where edges in a network are placed primarily according to some geometric rule but their connections are also governed by some smaller second order effect. 
It may only be possible to measure the smaller effect once we have accounted for the primary geometric one by assigning coordinates using Lorentzian MDS or a similar method.
We can see this effect clearly when the geometric embedding is one in real geographic space, such as in \cite{Expert2011} where accounting for geographic distance in phone-call data allows more accurate prediction of the second order effect of shared language.\newline

Other approaches to geometric embedding exist in the literature.
MDS is characterised by maintaining global separations between pairs.
Others, such as Isomap\cite{Tenenbaum2000, Roweis2000} maintain shortest paths between pairs linked by local interactions.
That approach may not be as appropriate in pseudo-Riemannian geometries.
Firstly, the shortest path locally may not correspond to the geodesic distance like it does in Euclidean space; as discussed above in graphs embedded in Minkowski space it is the longest path which corresponds to the geodesic.
Secondly, the idea of local neighbours is less clearly defined if there are different types of separation, or if, as is the case for Minkowski space, the number of nearest neighbours diverges.
Another class of embedding approaches are probabalistic, such as Stochastic Neighbour Embedding \cite{Hinton2002}.
Although it is beyond the scope of this work, we do not see why such approaches could not be adapted to pseudo-Riemannian manifolds, and the ability to use a mixture of separated images for the same object may prove very useful.\newline

Finally we note that inserting a metric signature into the equations for classical MDS allows it to be used on any metric signature, even though we have focused only on the Lorentzian signature here.
To our knowledge this pseudo-Riemannian output is a new development, although some kind of manifold learning techniques exist which can take pseudo-Riemannian manifolds as their input \cite{Liu2010, Liu2010a}.
We note that when performing the Lorentzian network MDS algorithm we often find multiple negative eigenvalues, suggesting that embeddings in spaces with more than one timelike dimension is also possible, as are potential embeddings into Lorentzian manifolds other than Minkowski space, incorporating curvature or preferred directions.
\newpage

\newpage

\appendix
\section{Derivation of Minkowski MDS}
Following closely the derivation for Euclidean MDS in \cite{Cox2000} we begin with the timelike (negative) and spacelike (positive) square distances between each pair of points, and wish to derive the inner product matrix $\Bmat$, where $B_{ij} = M (\mathbf{x}_i, \mathbf{x}_j)$. 
We will denote the Minkowski separation between $i$ and $j$ as $S_{ij}= M (\mathbf{x}_i-\mathbf{x}_j, \mathbf{x}_i-\mathbf{x}_j)$. 

As usual we first fix the coordinates centre of mass, placing it at the origin, such that $\sum_{i=1}^{n}x_{i\mu} = 0$ for each $\mu=0, 1, 2, ...d$ in a $d+1$-dimensional space. 
Remembering that $M (\mathbf{x}_i, \mathbf{x}_j) = \mathbf{x}_i^\Trans \Gmat \mathbf{x}_j$ with $\Gmat$ a diagonal $n \times n$ matrix with $-1$ in the first row and $1$ in every other row (the Minkowski metric) we have:
\begin{align*}
S_{ij} & = \mathbf{x}_i^\Trans \Gmat \mathbf{x}_i
+ \mathbf{x}_j^\Trans \Gmat \mathbf{x}_j 
- 2 \mathbf{x}_i^\Trans \Gmat \mathbf{x}_j  \\
\frac{1}{n} \sum_{i=1}^{n} S_{ij} & =
\mathbf{x}_j^\Trans \Gmat \mathbf{x}_j 
+ \frac{1}{n} \sum_{i=1}^{n} \mathbf{x}_i^\Trans \Gmat \mathbf{x}_i
- \frac{2}{n} \sum_{i=1}^{n} \mathbf{x}_i^\Trans \Gmat \mathbf{x}_j \\
\frac{1}{n} \sum_{i=1}^{n} S_{ij} & =
\mathbf{x}_j^\Trans \Gmat \mathbf{x}_j 
+ \frac{1}{n} \sum_{i=1}^{n} \mathbf{x}_i^\Trans \Gmat \mathbf{x}_i
- \frac{2}{n} \sum_{i=1}^{n} \sum_{\mu=0}^d x_{i\mu} G_{\mu\mu} x_{j\mu} \\
\end{align*}
The last term on the right vanishing if the order of the sums is reversed due to the condition we stipulated above. Doing the same summing over $s$ as well gives:
\begin{align*}
\frac{1}{n} \sum_{i=1}^{n} S_{ij} & =
\mathbf{x}_j^\Trans \Gmat \mathbf{x}_j 
+ \frac{1}{n} \sum_{i=1}^{n} \mathbf{x}_i^\Trans \Gmat \mathbf{x}_i \\
\frac{1}{n} \sum_{j=1}^{n} S_{ij} & =
\mathbf{x}_i^\Trans \Gmat \mathbf{x}_i 
+ \frac{1}{n} \sum_{j=1}^{n} \mathbf{x}_j^\Trans \Gmat \mathbf{x}_j \\
\frac{1}{n^2} \sum_{i=1}^{n} \sum_{j=1}^{n} S_{ij} & =
\frac{2}{n} \sum_{i=1}^{n} \mathbf{x}_i^\Trans \Gmat \mathbf{x}_i\\
\end{align*}
We can now write down each element in the matrix $B$ we are trying to calculate.
\begin{align*}
B_{ij} & = \mathbf{x}_i^\Trans \Gmat \mathbf{x}_j \\
B_{ij} & = -\frac{1}{2} \left[ S_{ij}
- \mathbf{x}_i^\Trans \Gmat \mathbf{x}_i 
- \mathbf{x}_j^\Trans \Gmat \mathbf{x}_j \right] \\
B_{ij} & = -\frac{1}{2} \left[ S_{ij}
- \frac{1}{n} \sum_{i=1}^{n} S_{ij} 
+ \frac{1}{n} \sum_{i=1}^n \mathbf{x}_i^\Trans \Gmat \mathbf{x}_i 
- \frac{1}{n} \sum_{j=1}^{n} S_{ij} 
+ \frac{1}{n} \sum_{j=1}^n \mathbf{x}_j^\Trans \Gmat \mathbf{x}_j
\right] \\
B_{ij} & = -\frac{1}{2} \left[ S_{ij}
- \frac{1}{n} \sum_{i=1}^{n} S_{ij} - \frac{1}{n} \sum_{j=1}^{n} S_{ij}
+ \frac{1}{n^2} \sum_{i=1}^{n} \sum_{j=1}^{n} S_{ij} \right]
\end{align*}
Which gives us the the matrix $\Bmat$ from the distances.

In the Euclidean case all of the eigenvalues of $\Bmat$ are positive (or zero).
In general, the signs of eigenvalues of the matrix $\Bmat$ will follow the signature of the metric $\Gmat$.
For any metric, $\Bmat$ is symmetric and so can can be decomposed into orthogonal eigenvectors, $\Umat$ and eigenvalues $\Sigmamat$.
\begin{align*}
\Bmat = \Umat \Sigmamat \Umat^\Trans
\end{align*}
We aim to find a solution to the equation $\Bmat = \Xmat \Gmat \Xmat^\Trans$.
Trying $\Xmat = \Umat \Dmat$ where $\Dmat$ is some real diagonal matrix gives
\begin{align*}
\Umat \Sigmamat \Umat^\Trans & = \Xmat \Gmat \Xmat^\Trans \\
\Umat \Sigmamat \Umat^\Trans & = \Umat \Dmat \Gmat \Dmat \Umat^\Trans \\
\Sigmamat & = \Dmat \Gmat \Dmat
\end{align*}
Assuming that the metric $\Gmat$ is diagonal we then have that $\Sigma_i = D_i^2 G_i$ and since $\Dmat$ is real, the signs of the elements of $\Sigmamat$ must equal those of $\Gmat$.
\end{document}